%% file: HiddenMarkovEntropyRate.tex
\documentclass[a4paper,UKenglish]{lipics-v2016}
\usepackage{microtype}%if unwanted, comment out or use option "draft"

\title{Explicit Formulas on Renyi Entropy of Hidden Markov Models}
\titlerunning{Explicit Formulas on Renyi Entropy of Hidden Markov Models} %optional, in case that the title is too long; the running title should fit into the top page column

\author[1]{Joachim Breitner}
\affil[1]{
University of Pennsylvania \\
\href{mailto:joachim@cis.upenn.edu}{joachim@cis.upenn.edu}}

\author[2]{Maciej Skorski}
\affil[2]{IST Austria\\
  \href{mailto:maciej.skorski@gmail.com}{maciej.skorski@gmail.com}}
\authorrunning{J.\,Breitner and M.\,Skorski} %mandatory. First: Use abbreviated first/middle names. Second (only in severe cases): Use first author plus 'et. al.'

\Copyright{Joachim Breitner and Maciej Skorski}%mandatory, please use full first names. LIPIcs license is "CC-BY";  http://creativecommons.org/licenses/by/3.0/

\subjclass{G.1.2 Approximation, G.3 Statistical computing}% mandatory: Please choose ACM 1998 classifications from http://www.acm.org/about/class/ccs98-html . E.g., cite as "F.1.1 Models of Computation". 
\keywords{hidden Markov models, Renyi entropy, entropy rate, leakage modelling}% mandatory: Please provide 1-5 keywords
\EventEditors{John Q. Open and Joan R. Acces}
\EventNoEds{2}
\EventLongTitle{42nd Conference on Very Important Topics (CVIT 2016)}
\EventShortTitle{Submitted}
\EventAcronym{Submitted}
\EventYear{2016}
\EventDate{December 24--27, 2016}
\EventLocation{Little Whinging, United Kingdom}
\EventLogo{}
\SeriesVolume{42}
\ArticleNo{23}
\usepackage[utf8]{inputenc}

\usepackage{tikz}
\usetikzlibrary{shapes,decorations, patterns, calc, backgrounds,arrows,decorations.markings}
\usepackage{pgfplots}
\pgfplotsset{compat=1.10}
\usepgfplotslibrary{fillbetween}

\usepackage{amsmath, amssymb, amsthm}

\usepackage{enumerate}

\newtheorem{corollary}{Corollary}

\newtheorem{lemma}{Lemma}

\usepackage[linesnumbered,algoruled,boxed,boxruled,lined]{algorithm2e}
\SetCommentSty{medium}

\usepackage{multirow}

\usepackage{caption} 
\usepackage[backend=bibtex,style=alphabetic,maxnames=5,minnames=5,maxbibnames=99,firstinits=true,
doi=false,isbn=false,url=true
]{biblatex}
\AtEveryBibitem{%
    \clearfield{pages}
}

\usepackage{hyperref}
\hypersetup{
  colorlinks=true,
  pdfmenubar=false,
  pdfstartview={FitH},
  linktoc=all,
  urlcolor=blue,
  linkcolor={blue}
}
\usepackage[nameinlink, capitalize]{cleveref}

\usepackage[nameinlink]{cleveref}
\crefname{table}{table}{tables}
\Crefname{table}{Table}{Tables}
\crefname{figure}{figure}{figures}
\Crefname{figure}{Figure}{Figures}
\crefname{section}{section}{sections}
\Crefname{section}{Section}{Sections}
\Crefname{lemma}{Lemma}{Lemmas}
\crefname{claim}{claim}{claims}
\Crefname{claim}{Claim}{Claims}
\Crefname{question}{Question}{Questions}
\Crefname{appendix}{Appendix}{Appendices}

\newcommand{\cS}{\mathcal{S}}
\newcommand{\cC}{\mathcal{C}}

\newcommand{\cZ}{\mathcal{Z}}
\newcommand{\cX}{\mathcal{X}}

\newtheorem{claim}{Claim}

\usepackage{tikz}
\usetikzlibrary{patterns,positioning}

\title{Analytic Formulas for Renyi Entropy of Hidden Markov Models}

\begin{document}

\maketitle

\begin{abstract}
Determining entropy rates of stochastic processes is a fundamental and difficult problem, with closed-form solutions known only for specific cases. This paper pushes the state-of-the-art by 
solving the problem for Hidden Markov Models (HMMs) and Renyi entropies.

While the problem for Markov chains reduces to studying the growth of a matrix product, computations for HMMs involve \emph{products of random matrices}. As a result, this case is much harder and no explicit formulas have been known so far.
We show how to circumvent this issue for Renyi entropy of integer orders, reducing the problem again to a \emph{single matrix products} where the matrix is formed from transition and emission probabilities by means of tensor product.

To obtain results in the asymptotic setting,
we use a novel technique for determining the growth of non-negative matrix powers. The classical approach is
the Frobenius-Perron theory, but it requires positivity assumptions; we instead work directly with the spectral formula. As a consequence, our results
do not suffer from limitations such as irreducibility and aperiodicity. This improves our understanding of the entropy rate even for standard (unhidden) Markov chains.

A recently published side-channel attack against RSA was proven effective using our result, specialized to order 2.
% Bernstein explicitly asks people not to cite multi-author work as "Bernstein et at."

\end{abstract}

%\begin{keywords}
%\ Hidden Markov Moidels, Renyi Entropy, Entropy Rate
%\end{keywords}

\section{Introduction}

\subsection{Renyi Entropy of Stochastic Processeses}

The notion of Renyi entropy, introduced by Renyi in~\cite{renyi60entropy} as an extension of Shannon entropy, finds a number of applications accrross many disciplines including
coding theory~\cite{DBLP:journals/tit/Csiszar95}, unsupervised learning ~\cite{Xu:1998:EEI:929350,1223401}, anomaly detection~\cite{DBLP:conf/ica3pp/LiZYD09}, multiple source adaptation~\cite{DBLP:conf/uai/MansourMR09},
image processing~\cite{DBLP:conf/icip/MaHGM00,DBLP:journals/imst/NeemuchwalaHZC06,DBLP:journals/pr/SahooA04},
password guessing~\cite{DBLP:journals/tit/Arikan96,DBLP:journals/tit/PfisterS04,DBLP:journals/tit/HanawalS11},
randomness extraction~\cite{DBLP:conf/focs/ImpagliazzoZ89,DBLP:journals/tit/BennettBCM95},
testing random number generators~\cite{Knuth:1998:ACP:280635,DBLP:journals/joc/OorschotW99},
quantifying neural activity~\cite{Paninski2003},
%analysis of fractal structres~\cite{doi:10.1063/1.1427481},
or economy and finance ~\cite{10.2307/2344585,Jizba20122971}. In the finite sample regime, one defines the
 $\alpha$-th order Renyi entropy of a random variable $Z$ over a finite alphabet $\cZ$ as
\begin{align*}
H_{\alpha}(Z) = \frac{1}{1-\alpha}\log \sum_{z\in\cZ}P_Z(z)^{\alpha}
\end{align*}
whereas, for a stochastic source $Z=\{Z_i\}_{i=1}^{\infty}$ 
the quantity of interest is the entropy per output symbol in a 
\emph{finite-length realization} $Z^{n}_1 = Z_1,\ldots,Z_n$ and its limit
\begin{align*}
H_{\alpha}(Z) = \lim_{n\to\infty}\frac{H_{\alpha}(Z^{n}_1)}{n},
\end{align*}
called the \emph{entropy rate}. 
Finding explicit formulas for entropy rates or finite realizations
for general sources is is intracktable, and is remains non-trivial even for restricted classes of sources. So far, most general classes of sources
with known entropy formulas are \emph{Markov chains}, with the asymptotic analysis given~\cite{DBLP:journals/tit/RachedAC01} 
and the finite-length regimes studied recently in~\cite{DBLP:conf/isit/KamathV16}. In this paper the focus
is on \emph{Hidden Markov Models}, which are Markov chains observed through a noisy memoryless channel. 
While studying Renyi entropy is justified on its own right,
investigating HMM seems to be particularly important 
because they are widely regarded as very powerful models for sequential data, for example in natural language processing or bioinformatics.

\subsection{Summary of Our Results and Related Works}

We show how to \emph{explictly} compute the Renyi entropy of HMMs over finite alphabets in both finite-length and asymptotic regimes.
This problem has been open so far; the recent work~\cite{8007074} discuss only some convergence properties with no explicit formulas and under further restrictions.

For Shannon entropy the problem has been found
 hard and solvable only for specific cases, being related to an intracktable task in random matrix products - finding top Lapunov exponents~\cite{DBLP:journals/tcs/JacquetSS08}. Our main finding is that 
 calculating the Renyi entropy in finite-length regimes can be reduced to (explicit) \emph{powering of substochastic matrices};
 in the asymptotic regime powers can be approximated by \emph{spectral analysis} which yields formulas on entropy rates. While this technique was also used for Markov chains~\cite{DBLP:journals/tit/RachedAC01}, the (straightforward) reduction cannot be exteneded to our case. Our reduction is based on studying \emph{collision probabilities} of independent copies of the process, with explicit formulas obtained by the use of \emph{tensor products}. This part requires the assumption that the entropy order is an integer bigger than 1,
 which is a minor limitation of our results given that most of applications of Renyi entropy use integer orders.
 Interestingly, we are able to \emph{remove positivity assumptions}
 used before~\cite{DBLP:journals/tit/RachedAC01,DBLP:conf/isit/KamathV16,DBLP:journals/tcs/JacquetSS08,8007074} when discussing entropy rates.
 
 \Cref{table:1} gives a summary of our results
 compared to related literature.%, explaining our techniques and the main claim the next paragraph.
 
\begin{table}[b]
\caption{Formulas on entropy of stochastic processes.}
\resizebox{0.99\textwidth}{!}{
\setlength{\tabcolsep}{0.3em}
{\renewcommand{\arraystretch}{1.3}
\begin{tabular}{|l|l|l|l|l|}
\hline
Authors & Model & Entropy & Technique & Model Limitations \\
\hline
\cite{DBLP:journals/tit/RachedAC01,DBLP:conf/isit/KamathV16}
& Markov & Renyi & matrix powering & positivity assumptions \\
\hline
\cite{DBLP:journals/tcs/JacquetSS08}
& HMM (binary) & Shannon & random matrix products & positivity assumptions \\
\hline
\cite{8007074}
& HMM  & Renyi & Markov-approximations &  \parbox{4cm}{positivity assumptions \\ no explicit formula} \\
\hline
\textbf{this paper}
& HMM & Renyi & tensoring + matrix powering & none \\
\hline
\end{tabular}
}
}
\label{table:1}
\end{table} 

\subsection{Our Result and Techniques}

The problem of computing the Renyi entropy of a process $Z = \{Z_i\}_{i=1}^{\infty}$ over an alphabet $\cZ$ boils down to computing the empth{collision probability}
\begin{align}\label{eq:collision_pr}
\mathsf{CP}_{\alpha}(Z_1^{n}) = \sum_{z_1^n\in\cZ^n}p(z_1^n)^{\alpha}.
\end{align}

If $Z$ is a Markov chain, then we can factorize
$p(z_1^n) = p(z_i) \cdot \prod_{i=2}^{n}p(z_i|z_{i-1})$ 
which can ce computed as a \emph{product of one matrix} because
$p(z_i|z_{i-1}) = M(z_i;z_{i-1})$ where $M$ is the state transition matrix, not dependent on $i$. Thus \eqref{eq:collision_pr} depends
on \emph{matrix products} of the $\alpha$-\emph{entrywise} power of $M$, 
denoted by $M^{\diamond \alpha}$. Matrix powers, under extra positivity assumptions, can be approximated by the Perron-Frobenius theory~\cite{journals/siamrev/MacCluer00}. It follows that the asymptotic behavior is controled by the biggest eigenvalue of $M^{\diamond \alpha}$. In particular for large $n$ we have~\cite{DBLP:journals/tit/RachedAC01}
\begin{align}\label{eq:collision_pr_approx}
\mathsf{CP}_{\alpha}(Z_1^{n})& = \Theta(1)\cdot
\rho(M^{\diamond \alpha})^{n} \\
 \frac{ H_{\alpha}(Z_1^n)}{n}
& =  \frac{1}{1-\alpha}\log \rho(M^{\diamond \alpha})^{n}\cdot \big(1+o(1)\big),\quad n\to\infty.
\end{align}

For Hidden Markov Models, which are observations of some Markov process
$\{X_i\}_{i=1}^{\infty}$, this approach fails. This is because
the factorization $p(z_1^n)^{\alpha} = p(z_1)^{\alpha}\cdot \prod_{i=2}^{n}p(z_i|z_{i-1})^{\alpha}$ 
boils down to \emph{random matrix products}, as
the transition from $z_{i-1}$ to $z_i$ depends the hidden states $x_{i-1},x_{i}$ with changes following a random process.
The theory of asymptotic properties of random matrix products
is not only fairly involved buts so far insufficient for the problem at hand. We have explicit results for products of stochastic matrices~\cite{DBLP:conf/allerton/BajovicXS12}; however our matrices
$p(z_i|z_{i-1})^{\alpha}$ are sub-stochastic becuase of the $\alpha$-power. A tempting alternative might be the standard factorization conditioned on hidden states $p(z_1^n) = p(z_1,x_1)\cdot\sum_{x_1^n}\prod_{i=2}^{n}p(z_i|x_{i})p(x_{i}|x_{i-1})$. However, while it can be computed recursively by dynamic programming, it is not compatible with matrix multiplication when raised to the power $\alpha$ (as opposed to the previous case).

From now, we assume that $Z_i$ is a hidden Markov process with the underlying Markov chain $X_i$, both on finite alphabets.

\subsubsection{Closed formulas for Renyi entropies of HMMs}
To avoid getting into random matrix products or 
recurrences with no explicit solution, we change the approach
and observe that, in case of integer $\alpha$, we can see
\eqref{eq:collision_pr} as the probability that $\alpha$ independent finite length realizations collide. Thus, we are interested in the event
\begin{align*}
E_n = \Big\{\forall i=1,\ldots,n:\ Z^{(1)}_i = Z^{(2)}_i = \ldots = Z^{(\alpha)}_i  \Big\}
\end{align*}
where $Z^{(1)},\ldots,Z^{(\alpha)}$ are independent copies of $Z$.
The probability of this event can be evaluated recursively by dynamic programming, conditioned on hidden states.
 More precisely, denote
for shortness the tuples of random variables
$X'_i = \big( X^{(1)}_i,\ldots, X^{(\alpha)}_i\big) $
and  $Z'_i = \big( Z^{(1)}_i,\ldots, Z^{(\alpha)}_i\big)$.
The probability of going from $Z^{(j)}_i = z^{(j)}_i$
to $Z^{(j)}_{i+1} = z^{(j)}_{i+1}$ is fully explained by the hidden states
$x^{(j)}_i$ and $x^{(j)}_{i+1}$. Namely
\begin{align}\label{eq:tensored_factoring}
\Pr[E_n]  = \sum_{(x'_i,z'_i)\in\cC} P_{X'_{i-1},Z'_{i-1} | X'_{i},Z'_{i}}\left((x'_i,z'_i | x'_{i-1},z'_{i-1}\right)
\end{align}
where the summation is restricted to tuples $x'_i,z'_i$ satisfies \emph{collision restrictions}
\begin{align}\label{eq:collision_set}
\cC = \{x^1,z^1\ldots,x^{\alpha},z^{\alpha} \in (\cX\times\cZ)^{\alpha}: 
z^1 = z^2 = \ldots = z^{\alpha} \}.
\end{align}
%We note that this formula cannot be immediately obtained  from the standard factorization used  to compute probabilities for HMMs. 
Each distribution
$P_{X'_{i+1},Z'_{i+1} | X'_{i},Z'_{i}}$ in \Cref{eq:tensored_factoring} is given by a fixed matrix, being the transition matrix of the process 
$\{X'_i,Z'_i\}_i$ which is Markov (once we have revealed hidden states, see \Cref{sec:hidden_aux}).
Denoting the transition matrix of $\{X_i,Z_i\}_i$ by $M$, we can find
the matrix of $\{X'_i,Z'_i\}_i$ as the $\alpha$-fold \emph{Kronecker tensor product} of $M$, denoted by $M^{\otimes \alpha}$, which is a matrix
with rows and columns in 
the $\alpha$-fold cartesian product of $\cX\times\cZ$; we call it 
the \emph{tensored matrix} of $M$.
Let $M^{\otimes \alpha}_{\cC}$ be its submatrix restricted to rows and columns satisfying the restriction $\cC$ in \Cref{eq:collision_set}, refereed to as
the \emph{restricted tensored matrix}, and 
let $\left(P_{X'_1,Z'_1}\right)_{\cC}$ be the restriction of the probability vector of $X',Z'$ to 
indices from $\cC$.

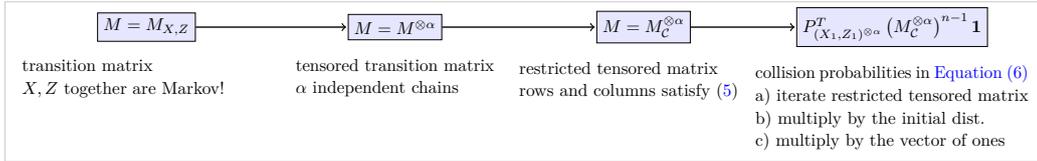
\begin{figure}[h!]
\resizebox{0.99\textwidth}{!}{
\fcolorbox{gray!40}{white}{
\begin{tikzpicture}
\node (a) at (0,0) [draw, rectangle, fill = blue!10] {$M = M_{X,Z}$};
\node[below =0.25cm  of a] {\parbox{5cm}{transition matrix\\$X,Z$ together are Markov!}};
\node (b) at (5,0) [draw, rectangle, fill= blue!10] {$M = M^{\otimes \alpha}$};\node[below =0.25cm  of b] {\parbox{4cm}{tensored transition matrix \\ $\alpha$ independent chains}};
\node (c) at (10,0) [draw, rectangle, fill = blue!10] {$M = M^{\otimes \alpha}_{
\cC}$};
\node[below =0.25cm  of c] {\parbox{5cm}{restricted tensored matrix\\ rows and columns satisfy \eqref{eq:collision_set} }};
\node (d) at (15,0) [draw, rectangle, fill = blue!10] {$ P_{(X_1,Z_1)^{\otimes \alpha}}^T \left(M^{\otimes \alpha}_{
\cC}\right)^{n-1} \mathbf{1}$};
\node[below =0.25cm  of d] {\parbox{5.5cm}{collision probabilities in \Cref{eq:entropy_finite_length}\\
a) iterate restricted tensored matrix\\
b) multiply by the initial dist. \\
c) multiply by the vector of ones}};
\draw[->,shorten >=1pt, thick] (a)--(b);
\draw[->,shorten >=1pt, thick] (b)--(c);
\draw[->,shorten >=1pt, thick] (c)--(d);
\end{tikzpicture}
}
}
\caption{The workflow of our framework for computing integer-order Renyi entropies.}
\label{fig:workflow}
\end{figure}

Then \Cref{eq:tensored_factoring} can be expressed in a compact form in the following way (multiplications are understood as matrix/vector multiplications)
\begin{theorem}[Renyi entropy in finite-length regimes]\label{thm:finite_length}
The entropy of finite-length realizations of $Z$ can be computed with 
powers of the restricted tensored product of $M$ as follows
\begin{align}\label{eq:entropy_finite_length}
H_{\alpha}(Z^n_1) = \frac{1}{1-\alpha}\log\left( 
\left(P_{X'_1,Z'_1}\right)_{\cC}^T\cdot \left( M^{\otimes \alpha}_{\cC}\right)^{n-1} \cdot \mathbf{1} \right)
\end{align} 
\end{theorem}
\begin{remark}[Explicitly computing the base matrix]
The matrix $M$ can be computed from emission and transition probabilities, respectively $p(z_i|x_i)$ and $p(x_i|x_{i-1})$ (see 
\Cref{sec:hidden_aux}). 
\end{remark}
Explicitly computing entropies can be illustrated with a diagram in \Cref{fig:workflow}.

The proof of \Cref{thm:finite_length} appears in 
\Cref{proof:finite_length}.

\subsubsection{Explicit Renyi entropy rates for HMMs without positivity}

To approximate the iterated powers of non-negative matrices
in \eqref{eq:entropy_finite_length} we can use the classical Perron-Frebenius theory. A drawback is, however, that this requires 
\emph{positivity assumptions} on the matrix - for example, that it has a strictly positive power. Phrased in terms of the stochastic process, this means that results would suffer from being only applicable to
to matrices with \emph{irreducible and aperiodic} supporting graphs.
In principle, one can decompose the matrix into components obeying positivity assumptions (for example the canonical decomposition into irreducible parts), and apply the Perron-Froenius theory separately. However, handling peridocity or even merging results from individual components is not immediate. One such counter-intuitive case is discussed in \Cref{remark:carefull}.

We give an elegant solution to a more general problem - the growth of certain \emph{pseudonorms} of matrix powers. Namely, for any non-negative matrix $A$ and a
non-negative vector $u$ we determine the rate of growth
of $u^T\cdot A^n\cdot \mathbf{1}$ with $n$ (which in particular fits
\Cref{eq:entropy_finite_length}).
The mapping $B \to u^T\cdot |B| \cdot \mathbf{1}$, where $|B|$ is the element-wise application of the absolute value, is a weighted sum of absolute elements in $B$ with non-negative coefficients and thus a pseudonorm. Our problem reduces to estimating 
how powers of $A$ grow under this pseudonorm, which we handle by using  the \emph{spectral formula}. This result, stated later in \Cref{lemma:matrix_power_growth} and of independent interest, allows us to compute the rate of any hidden Markov process.
\begin{theorem}[Renyi entropy rates]\label{thm:Renyi_rates}
Let $\rho_i$ for $i\in I$ be the spectral radius of matrices corresponding to the irreducible components of $M^{\otimes \alpha}_{\cC}$. Then the entropy rate is given by the formula
\begin{align*}
H_{\alpha}(Z) = \frac{1}{1-\alpha}\log\Big( \max_{i\in I^{+}} \rho_i \Big)
\end{align*}
where the set $I^{+}$ of 'reachable' components is defined as 
all components that can be reached in the associated graph 
of the matrix $M^{\otimes \alpha}_{\cC}$ from tuples in $\cC$ having positive probability under the initial distribution $P_{X',Z'}$.
\end{theorem}

\begin{remark}[Positivity assumptions removed]
Note that the result depends on the initial distribution
and the dominant eigenvalue, but the matrix can be arbitrary.
\end{remark}

The proof of \Cref{thm:Renyi_rates} appears in 
\Cref{sec:Renyi_rates}.

\subsubsection{Key lemma: growth of non-negative matrix powers}

Below we abstract our main technical ingredient - the lemma
giving the growth rate of matrix powers under certain 'pseudonorms'. 
Since more limited results of this form have found applications in theory of random matrices~\cite{DBLP:conf/allerton/BajovicXS12} 
and previous works on entropy of HMMs~\cite{DBLP:journals/tit/RachedAC01,DBLP:journals/tcs/JacquetSS08}
 we belive it may be of independent interest.

\begin{lemma}[Weighted element sum of matrix powers]\label{lemma:matrix_power_growth}
Let $A$ be a non-negative matrix of size $m\times m$ and $u$ be a non-negative vector of length $m$. Let $I^{+}$ contain all $i\in\{1,\ldots,m\}$ with
\begin{align}\label{eq:assump_1}
    u^T \sum_{k=0}^{\infty} A^k e_i  > 0.
\end{align}
Let $A^{+}$ be the submatrix of $A$ with rows and columns $I^{+}$. Then
\begin{align}\label{eq:pseudo_Gelfand}
    \lim_{n\to\infty}\big( u^T A^n\mathbf{1}\big)^{\frac 1 n} = \rho(A^{+}).
\end{align}
\end{lemma}
\begin{remark}[Simplification by the associated graph and irreducible decomposition]
The description of $A^{+}$ can be simplified slightly by using the  associated graph and the irreducible components. See the proof of \Cref{thm:Renyi_rates} for details.
\end{remark}
The proof combines the canonical decomposition, sandwiching argument and Gelfand's formula. We give it in \Cref{proof:matrix_power_growth}.
A very special case when $A$ is irreducible and $u$ is positive follows from 
\cite{marcus1962}, and an easy proof for any matrix $A$ and positive $u$ can be found in \cite{DBLP:conf/allerton/BajovicXS12}. Our result is more general as we consider non-negative weights.

\subsection{Examples and Applications}\label{sec:apps}

\subsubsection{Example: tensoring and restricting step by step}

\input{fig-markovgraph.tex}

Consider the Markov chain $X$ in \Cref{fig:markovgraph}. Its hidden states are $\{1,2,3\}$, of which the observer cannot distinguish state 1 and 3, as indicated by color. We assume the starting distribution $X_1$ to be $\frac{1}{3}$ on every state. The transition matrix is
\begin{align*}
M = \begin{bmatrix}
0.9 & 0.1 &   0 \\
  0 & 0.4 & 0.6 \\
  0 & 0.6 & 0.4
 \end{bmatrix}.
\end{align*}

To calculate its collision entropy, we take the second tensor, and obtain the graph on the right of the picture. Intuitively, this corresponds models two independent copies of $X$.   The transition matrix on the states $\{11,12,13,21,22,23,31,32,33\}$ is
\begin{align*}
M^{\otimes 2} = \begin{bmatrix}
% 11     12     13     21     22     23     31     32    33
0.81 & 0.09 &      & 0.09 & 0.01 \\
     & 0.36 & 0.54 &      & 0.04 & 0.06 \\
     & 0.54 & 0.36 &      & 0.06 & 0.04 \\
     &      &      & 0.36 & 0.04 &      & 0.54 & 0.06 \\
     &      &      &      & 0.16 & 0.24 &      & 0.24 & 0.36 \\
     &      &      &      & 0.24 & 0.16 &      & 0.36 & 0.24 \\
     &      &      & 0.54 & 0.06 &      & 0.36 & 0.04 \\
     &      &      &      & 0.24 & 0.36 &      & 0.16 & 0.24 \\
     &      &      &      & 0.36 & 0.24 &      & 0.24 & 0.16 \\
\end{bmatrix}.
\end{align*}
But not all of these states are in $\cC = \{11,13,31,33,22\}$, so we consider the restriction to $\cC$, i.e.\@
\begin{align*}
M^{\otimes 2}_{\cC} = \begin{bmatrix}
% 11    13    22    31   33
0.81 &     & 0.01 \\
     &0.36 & 0.06 \\
     &     & 0.16 &     &0.36 \\
     &     & 0.06 &0.36  \\
     &     & 0.36 &     &0.16 \\
\end{bmatrix}.
\end{align*}
The intuition here is that we only care about executions where the observed behavior of the two independent copies of $X$ is indistinguishable.

To investigate the asymptotic behavior, 
we find that largest eigenvalues of two irreducible components
are $\rho_1 = 0.52$ and $\rho_2 = 0.81$ (the eigenvectors $e_1 = (0, 0, 1, 0, 1)$ and
$e_2 = (0.96666, 0, 0.02145, 0, 0.01188)$ describe a distribution that is stationary under the condition that we continue to observe the same output; the eigenvalue is the probability of continued collision)

%we calculate the positive eigenvectors, which are are
%$e_1 = (0, 0, 1, 0, 1)$ with eigenvalue $\rho_1 =0.52$ and
%$e_2 = (0.96666, 0, 0.02145, 0, 0.01188)$ with eigenvalue $\rho_2 = 0.81$. These eigenvectors describe a distribution that is stationary under the condition that we continue to observe the same output; the eigenvalue is the probability of continued collision.
Since $\rho_2$ is larger, it is the asymptotically relevant, and according to \Cref{thm:main} we have $H_2(X) = -\log(\rho_2) = 0.304$.

\subsubsection{Example: (lack of) relation to stationary distribution and recurrent states}\label{remark:carefull}

Consider again the transition matrix $M$ from the previous example, with all states being visible (so that $Z=X$). This demonstrates somewhat counter-intuitive behavior. 
The Markov chain converges to a stationary distribution where the first state has probability zero. In fact, this state is not recurrent:
with probability 1, the chain visits it only a finite number of times because $\sum_{k} M^{k}_{11} < \infty$ (the generic test for recurrent states). 

Intuitively, such a state should be negligible in the asymptotic entropy analysis.
However the opposite happens: for Renyi entropy of order $\alpha=2$ and any finite length $n$ the first state contributes most to the matrix powering ($0.81^{n}$ as opposed to $\lambda^{n}$ where $|\lambda| < 0.81$ contributed by the component formed by the second and third state). Thus in the asymptotic setting
the entropy depends only on the first state.
\begin{corollary}
The Renyi entropy rate is \emph{not related to  limiting distributions or recurrent/transient properties of states}.
\end{corollary}

\subsubsection{Noiseless Observations}

We consider study the hidden Markov model where 
the state chain is observed through \emph{noiseless measurements}. More precisely, for a deterministic mapping $T:\cX\rightarrow \cZ$ the observed (hidden) chain $Z_i$ is given by mapping the base Markov chain: 
$Z_i = T(X_i)$. While
this is less general than our result in \Cref{thm:Renyi_rates}, this particular case leads to a \emph{very sparse tensored matrix} and more compact
formula is possible (independent on the dimension of $Z$).

\begin{theorem}[More compact formula for noisyless case]\label{thm:main}
Let $X_i$ be as above with the transition matrix $M$. Let $M_{\cC}^{\otimes \alpha}$ be the $\alpha$-fold Kronecker tensor product of $M$ restricted to the tuples of indices $s=(s_1,\ldots,s_{\alpha})$ such that $T(s_1)=T(s_2)=\ldots =T(s_{\alpha})$. Then for any integer $\alpha>1$:
\begin{itemize}
\item the entropy rate of $Z_i$ measured by Renyi entropy of order $\alpha$ is given by
\Cref{thm:Renyi_rates} applied to $M_{\cC}^{\otimes \alpha}$ as above and the initial distribution being $P_{X_1^{\otimes \alpha}}$ ($\alpha$-fold product of $X_1$).
\item moreover, if $M$ is irreducible and aperiodic, the entropy rate is given by a simpler formula
\begin{align*}
    H_{\alpha}(\{Z_i\}_i) =  \frac{1}{\alpha-1}\cdot \rho( M_{\cC}^{\otimes \alpha} )
\end{align*}
where $\rho(\cdot)$ denotes the spectral radius.
\end{itemize}
\end{theorem}
The proof appears in \Cref{proof:main}.

\subsubsection{Modelling Side Channel Leakage}\label{sec:leakage}

The motivation and first application for this work was the theoretical analysis of a side-channel attack against RSA encryption~\cite{cryptoeprint:2017:627}. By observing memory access timing, the attacker gains knowledge about what instructions the victim's encryption program is executing. In this particular case, as the victim performs the modular exponentiation necessary for encryption using a sliding-window square-and-multiply algorithm, the attacker learns the sequence of squares and multiplies performed.

The attacker tries to recover the secret key using an established \emph{search-and-prune} technique.
%~\cite{search-and-prune}. 
By using the leaked observation, he can prune the tree more aggressively. This attack is practical if the size of this search tree is linear in the size of the secret key.

The formula that bounds the size of the search tree from above turns out to depend directly on the collisions entropy of the stochastic process that models the leaked observations (assuming a random and uniformly distributed key). This make intuitive sense, as the tree is pruned if two independent copies of this process (the real one and the guessed one) no longer collide.
Concretely, the search tree is linear in the size of the secret key if the entropy rate is $H<0.5$.

By modelling the states of the square-and-multiply algorithm a Markov chain and the observation therefore as a HMM, we can apply \cref{thm:main} and obtain $H=0.545$, explaining why the attack is successful.

\subsubsection{Entropy rates for finite Markov chains}

If $T$ in \Cref{thm:main} is a one-to-one mapping, and $X_i$ is aperiodic and irreducible, then
the rate equals $\frac{1}{1-\alpha}\log\rho(M^{\diamond\alpha})$ where $M^{\diamond\alpha}$ is the $\alpha$-fold Hadamard product. This reproves the formula for Markov chains with no hidden states~\cite{DBLP:journals/tit/RachedAC01}. For an illustrative example and details see \Cref{sec:markov_rates}.

In fact, the more general part of \Cref{thm:main} holds with no positivity (aperiodicity and irreducibility) assumptions. 
Moreover, although it
uses the assumption that $\alpha$ is integer, for this case we can use directly \Cref{lemma:matrix_power_growth} in the analysis \cite{DBLP:journals/tit/RachedAC01}, instead of Perron-Frobenius theory. We thus extend the classical result to possibly periodic and reducible Markov chains
\begin{corollary}[Renyi entropy for any Markov chain]
Let $\alpha\not=1$ be a positive real number. Let $Z_i$ be a finite-alphabet Markov chain with a transition matrix $M$. Let $\rho_i$ for $i\in I$ be the spectral radius of all irreducible components of $M^{\diamond \alpha}$. Then the entropy rate is given by 
\begin{align*}
H_{\alpha}(Z) = \frac{1}{1-\alpha}\log\Big( \max_{i\in I^{+}} \rho_i \Big)
\end{align*}
where the maximum is taken over all 'positive' components $I^{+}$ that
are assigned positive mass under the initial distribution $X_1$.
\end{corollary}

\subsubsection{Binary Markov chains and Bernoulli noise}

If the chain outcomes are flipped with probability $\epsilon$ by a noisy channel, 
the rate (for order $\alpha>1$) changes by an $O(\epsilon)$ term. The exact expression up to $O(\epsilon^2)$ has been studied in~\cite{DBLP:journals/tcs/JacquetSS08}. 
We provide an alternative characterization of the entropy rate, as the root of an \emph{explicit polynomial of degree 8}. In particular, 
for any $\epsilon$ we can compute the exact value numerically, without asymptotic expressions. See \Cref{sec:binary_noise} for details.
For this application we assume that the transition matrix is positive, to apply perturbation theory.

\subsection{Algebraic equations for the entropy rate}

Our results imply that the Renyi entropy rate of integer order $\alpha>1$ is characterized by an algebraic equation.
\begin{corollary}
The Renyi entropy rate of integer order $\alpha>1$ of a hidden Markov process (with the base chain over a finite alphabet) is the absolute value of a root of some explicit polynomial.
\end{corollary}
This is interesting when compared to the Shannon entropy rate can be characterized by a more complicated functional equation~\cite{DBLP:journals/tit/LuoG09}.
As a consequence of this fact, one can derive exact Taylor approximations, efficient numerical approximations, or study perturbations (e.g. due to noise changes).

\section{Preliminaries}

We first discuss some notational conventions. 
For a process $Z=Z_1,Z_2,\ldots$ we define the finite realization of length $n$ as $Z^n_1 = Z_1,\ldots,Z_n$.
To simplify the notation, we use the standard convention that probabilities involving events of the form
$A_i=a_i,B_i=b_i$ are written with capital symbols omitted, that is $P(A_i = a_i) = p(a_i)$, $P(A_i=a_i|B_i=b_i) = p(a_i|b_i)$ and so on.
We identify the probability distribution of a random variable $S$ with values in (finite) $\cS$ with the vector with coordinates indexed by $\cS$.
For any vector $\mu$ or matrix $A$ indexed by $\cS$, by $\mu_{\cS'}$ respectively $A_{\cS'}$ we understand restrictions to indices from $\cS'$ (applies to $\cS'\subset \cS$). 
Single vectors are understood as columns; $y^T$ denotes the transposition of a vector or matrix $y$.
All logarithms are taken at base $2$.

\begin{definition}[Associated graph of a matrix]
For a non-negative matrix $A$ the \emph{associated graph} (or supporting graph) is the directed graph with all matrix indices $1,\ldots,m$ as nodes, and edges $i\to j$ if and only if $A_{i,j} > 0$ for all $i,j$. 
\end{definition}

\begin{definition}[Renyi Entropy~\cite{renyi60entropy}]
The Renyi entropy of order $\alpha$ of a discrete random variable $Z$ is defined as
\begin{align*}
    H_{\alpha}(Z) = \frac{1}{1-\alpha}\log \sum_{z} P_Z(z)^{\alpha}
\end{align*}
with the Shannon entropy $H_1$ understood as the limit $\alpha\to 1$
\begin{align*}
    H_{1}(Z) = -\sum_{z}P_Z(z)\log  P_Z(z)
\end{align*}
and the min-entropy $H_{\infty}$ being the limit $\alpha\to\infty$
\begin{align*}
    H_{\infty}(Z) = \min_{z}\log( 1/P_Z(z) ).
\end{align*}
\end{definition}

\begin{definition}[Entropy Rate]
The Renyi entropy rate, of order $\alpha$, of a discrete process $Z = \{Z_i\}_{i\geqslant 1}$ is defined as
\begin{align*}
    H_{\alpha}(Z) = \lim_{n\to\infty}\frac{1}{n}H_{\alpha}(Z_1,\ldots,Z_n)
\end{align*}
\end{definition}

\begin{definition}[Kronnecker Tensor Product]
\label{def:tensor_product}
For any two square matrices $A,B$ over $\cX\times \cX$, the Kronnecker product is a matrix $A\otimes B = C$ over $\cX^2\times\cX^2$ with the entries $C((i,j),(i',j')) = A(i,j)\cdot B(i',j')$
\end{definition}
The $\alpha$-fold tensor product of a matrix $A$ is denoted by $A^{\alpha}$. Sometimes for shortness we will also denote by $P_{Y^{\otimes \alpha}}$ the joint distribution of $\alpha$-independent copies of a distribution $P_Y$.
\begin{definition}[Hidden Markov Model]
The hidden Markov model consists of the base (hidden) chain $X_i$ and observations $Z_i$, for $i=1,2,\ldots$ such that the following two conditions are satisfied
\begin{enumerate}
\item Markov assumption: $P_{X_i|X_{i-1},\ldots,X_1} = P_{X_i|X_{i-1}}$
\item Output independence: $P_{Z_i|X_1,Z_1\ldots,X_i,Z_i\ldots,X_T,Z_T} = P_{Z_i|X_i}$.
\end{enumerate}
and the transition $P_{X_i|X_{i-1}}$ and emission $P_{Z_i|X_i}$ probabilities don't change with time $i$.
\end{definition}

\section{Main results}

\subsection{Proof of \Cref{thm:finite_length}}
\label{proof:finite_length}
\begin{proof}
For convenience, we assume that the processes $X$ and $Z$ are indexed starting from $i=0$.
For every $z\in \cZ$ we have
\begin{align*}
P_{Z^{n}}(z)^{\alpha} & =  p(z_0)^{\alpha} \prod_{i=1}^{n}p(z_i|z_{i-1})^{\alpha} \\
 & =  \left(\sum_{x_0,\ldots,x_n}p(z_0,x_0)\prod_{i=1}^{n}p(z_i,x_i|z_{i-1},x_{i-1})\right)^{\alpha}
\end{align*}
Defining, as in \Cref{eq:collision_set}
\begin{align*}
    \cC = \{(x^{1},z^{1},\ldots,x^{(\alpha)},z^{(\alpha)})\in (\cX\times\cZ)^{\alpha}: z^{(1)}=\ldots=z^{(\alpha)} \}
\end{align*}
we can write
\begin{align*}
\sum_{z}P_{Z^{n}}(z)^{\alpha} & =  \sum_{(x_{i},z_{i}) \in \cC}\prod_{i=1}^{n} 
\prod_{j=1}^{\alpha}p( z_i^{(j)},x_i^{(j)}|
z_{i-1}^{(j)},x_{i-1}^{(j)}  )
\prod_{j=1}^{\alpha} p(z_0^{(j)},x_0^{(j)})
\end{align*}
where $x_i = (x_i^{(1)},\ldots,x_i^{(\alpha)})$ and 
 $z_i = (z_i^{(1)},\ldots,z_i^{(\alpha)})$, and 
 for notational simplicity we use the isomorphism 
 $(\cX\times\cZ)^{\alpha}\cong \cX^{\alpha}\times \cZ^{\alpha}$ identifying
\begin{align*}
 (x_i^{(1)},z_i^{(1)},\ldots,x_i^{(\alpha)},z_i^{(\alpha)}) \cong ( (x_i^{(1)},\ldots,x_i^{(\alpha)}), (z_i^{(1)},\ldots,z_i^{(\alpha)}) ) =
  (x_i,z_i)
\end{align*}
This can be further simplified as follows:
let $M^{\otimes \alpha}_{\cC}$ be the $\alpha$-fold tensor product of the matrix
$M = p((z,x)|(z',x'))$ restricted to $\cC$, let
$\mu$ be the vector with elements
$\prod_{j=1}^{\alpha} p(x_0^{j},z_0^{j})$
over all choices $x_0,z_0 \in \cC$,
and let $\mathbf{1}$ be the vector of all ones of length
indexed by $\cC$. Then we have
\begin{align*}
\sum_{z}P_{Z^{n}}(z)^{\alpha} = \mu^T\cdot \left(M^{\otimes \alpha}_{\cC}\right)^{n}\cdot \mathbf{1}
\end{align*}
which together with the Renyi entropy definition implies
\Cref{eq:entropy_finite_length}.
\end{proof}

\subsection{Proof of \Cref{lemma:matrix_power_growth}}\label{proof:matrix_power_growth}

\begin{proof}
Decompose $A$ into irreducible communicating classes, reordering the states if necessary (the canonical decomposition for non-negative matrices\footnote{We say that $i$ communicates with $j$ if there is a path from $i$ to $j$ in the (directed) graph supporting $A$.}):
\begin{align*}
    A = \begin{bmatrix}
         A_1 & * & * & * \\
         0   & A_2 & * & * \\
         \ldots \\
         0 & 0 & 0 & A_d
    \end{bmatrix}
\end{align*}
and let $u_i$ be the corresponding parts of the vector $u$.
Let $I = \{i: u_i > 0\}$. Take any $i\not\in I$ 
Let $A'$ be the matrix obtained from $A$ by setting the $i$-th block row to zero. We have $u^TA = u^T A'$
and $A'A = A'A'$, thus $u^T A^n = u^T(A')^n$ for any $n$.
Therefore we may replace $A$ by $A'$ not affecting the limit.
Continuing this approach, we see that we can setting to zero all row blocks with indexes not in $I$. Denote the obtained matrix by $A'$ and its diagonal blocks by $A'_i$. We have
\begin{align*}
\left( u^T A^n\mathbf{1}\right)^{1/n} = \left( u^T (A')^n\mathbf{1}\right)^{1/n}
\end{align*}
Note that $u^T B\mathbf{1} \leqslant O(1) \|B\|_{1}$
where $\|\cdot\|_1$ is the matrix $\ell_1$-induced norm.
Therefore, by the Gelfand formula
\begin{align*}
\left( u^T (A')^n\mathbf{1}\right)^{1/n} = \rho(A')+o(1)
\end{align*}
Since $A$ and $A'$ are triangular we have 
\begin{align*}
\rho(A') = \max_{i}\rho(A'_i) = \max_{i\in I}\rho(A_i)
\end{align*}
where we used the fact that $A'_i = A_i$ when $i\in I$ and $A'_i=0$ otherwise. This discussion shows that
\begin{align*}
\left( u^T A^n\mathbf{1}\right)^{1/n} \leqslant \max_{i\in I}\rho(A_i) + o(1)
\end{align*}
Since $A^{+}$ is triangular with diagonal components $\{A_i\}_{i\in I}$
we 
\begin{align*}
\left( u^T A^n\mathbf{1}\right)^{1/n} \leqslant \rho(A^{+}) + o(1) 
\end{align*}
To prove the opposite direction, we simply note that
\begin{align*}
\left(    u^T A^n \mathbf{1}\right)^{1/n} \geqslant  
\left(  u_i^T A_i^n \mathbf{1}_i  \right)^{1/n}
\end{align*}
for every $i\in I$. If can we prove $\left(u_i^T A_i^n\mathbf{1}_i\right)^{1/n} = \rho(A_i)+o(1)$ for every $i$, we obtain the claimed bound $\left(    u^T A^n \mathbf{1}\right)^{1/n}\geqslant \max_{i}\rho(A_i) = \rho(A^{+})$.

Consider now the case of single irreducible component $A_i = A$. By scaling both sides, we may assume that $A$ is substochastic (rows sum up to numbers smaller than $1$). Then 
\begin{align*}
    d\cdot  u^TA^n\mathbf{1} = \sum_{k=0}^{d} (u^TA^k) A^{n-k}\mathbf{1}
\end{align*}
But we have $A\mathbf{1}\leqslant \mathbf{1}$ since $A$ is substochastic. Therefore
$A^{n}\mathbf{1}\leqslant A^{n-k}\mathbf{1} \leqslant A^{n-d}\mathbf{1}$ for $k=0,\ldots,d$ and
\begin{align*}
\left(\sum_{k=0}^{d} u^TA^k\right) A^{n}\mathbf{1} \leqslant    d\cdot  u^TA^n\mathbf{1} \leqslant \left(\sum_{k=0}^{d} u^TA^k\right) A^{n-d}\mathbf{1}
\end{align*}
Let $u^{+} = \sum_{k=0}^{d} uA^k$, by the assumptions we know that $u^{+}>0$.
Th above inequality can be written as
\begin{align*}
    \|A^n\| \leqslant d\cdot u^TA^n\mathbf{1} \leqslant \| A^{n-d}\|
\end{align*}
where $\|B\| =  (u^{+})^T B\mathbf{1}$ is a matrix norm. The Gelfand formula gives us
\begin{align*}
\rho(A)+o(1) \leqslant  \left( uA^n\mathbf{1}\right)^{1/n} \leqslant \left(\rho(A)+o(1)\right)^{1-d/n}
\end{align*}
and taking $n\to\infty$ finishes the proof.
\end{proof}

\subsection{Proof of \Cref{thm:Renyi_rates}}
\label{sec:Renyi_rates}

The result follows from \Cref{lemma:matrix_power_growth} combined with \Cref{thm:finite_length}, once we prove the alternative characterization of 'positive induces' in 
\Cref{lemma:matrix_power_growth}.

\begin{claim}
Let $A$ be a non-negative matrix and $u$ be a non-negative vector. 
Let $I_1,\ldots,I_d$ be the subsets of indices corresponding to $d$ irreducible classes in the canonical decomposition of $A$. 
Then $\sum_{k} u A^k e_i > 0$
if and only if the associated graph of $A$ connects some $j$ such that $u_j > 0$ and $i$.
\end{claim}

This follows immediately from the properties of adjacency matrix -namely  
$A^k_{i,j} > 0$ if there is a path from $i$ to  $j$ of length $k$.

Translated to the setting of \Cref{thm:Renyi_rates},
this means all irreducible classes of indices that can be reached from a point with positive measure $(P_{X',Z'})_{\cC}$.

\section{Conclusion}

We have shown how to analytically compute
the Renyi entropy of hidden Markov processes, when the entropy order is an integer bigger than 1. The main technical contribution is an auxiliary result on \emph{pseudonorms of iterated matrices}, that allow us (for the first time) to \emph{get rid of positivity assumptions} when computing Renyi entropy rates. Some problems we leave for future work are:
\begin{itemize}
\item Is it possible to analytically compute
rates for non-integer $\alpha$?
\item The rates of smooth Renyi entropy.
\item The speed of the convergence towards the entropy rate. 
\item Application of perturbation theory to handle small leakages.
\end{itemize}

%We also note that our auxiliary fact on matrix pseudonorms may be useful to derive results in substochastic matrices 

% or

\printbibliography

\appendix

%Since the canonical decomposition is a block-diagonal matrix (possibly after rearranging indices),
%we easily conclude that the formula is the maximal spectral radius over those irreducible components which have an index with the positive %coefficient in $u$.

\section{Proof of \Cref{thm:main}}\label{proof:main}
We start with the proof of the second part. Let $Z^{n} = Z_0,\ldots,Z_n$.
Denote $p(z_i|z_{i-1}) = \Pr[Z_i=z_i | Z_{i-1}=z_{i-1}]$,
 $p(z_i,x_i|z_{i-1},x_{i-1}) = \Pr[Z_i=z_i,X_i=x_i | Z_{i-1}=z_{i-1},X_{i-1}=x_{i-1}]$, and
 $p(z_i,x_i) = \Pr[Z_i=z_i,X_i=x_i]$, $p(z_i) = \Pr[Z_i=z_i]$.
We can assume that the probability of $X_0$ is strictly positive.
Indeed, it is enough to consider the entropy rate of a process $Z_m,Z_{m+1},\ldots$ for $m$ large enough. 
Then the new starting distribution is given by $q'=P^{m}q_0$.
Since for $m$ large enough we have $P^m > 0$, we obtain
that all entries of $q'$ are positive.

By the Markov property, for any fixed $z=(z_1,\ldots,z_n)$ we have
\begin{align*}
P_{Z^{n}}(z)^{\alpha} & =  p(z_0)^{\alpha}\prod_{i=1}^{n}p(z_i|z_{i-1})^{\alpha} \\
 & =  \left(\sum_{x_0,\ldots,x_n}p(z_0,x_0)\prod_{i=1}^{n}p(z_i,x_i|z_{i-1},x_{i-1})\right)^{\alpha} \\
  & =  \left(\sum_{x_0,\ldots,x_n}p(z_0,x_0)\prod_{i=1}^{n}p(z_i|x_i)p(x_i|x_{i-1})\right)^{\alpha} \\
\end{align*}
(see for example~\cite{Chen2015} for more general derivations). 
For noiseless measurements this further simplifies to
\begin{align*}
 P_{Z^{n}}(z)^{\alpha} &  = \left(\sum_{x_i\in T^{-1}( z_i)}p(x_0)\prod_{i=1}^{n}p(x_i|x_{i-1})\right)^{\alpha}
\end{align*}
Let $P$ be the transition matrix of the state chain $\{X_i\}_i$. For any integer $k$, consider the \emph{$k$-fold Kronnecker product} of $P$
\begin{align*}
    p^{\otimes k}(s|s') = \left[ \prod_{j=1}^{k} p(s_j|s'_j) \right]_{s,s'}
\end{align*}
where $s=(s_{1},\ldots,s_{k})$
and $s'=(s'_{1},\ldots,s'_{k})$. For integer $\alpha$ and any fixed $(z_i)_i$ by the multinomial formula we obtain
\begin{align*}
P_{Z^{n}}(z)^{\alpha} 
& = 
  \sum_{x^{(i)}}
   \prod_{j=1}^{\alpha}p(x^{0}_j) \prod_{i=1}^{n} p^{\otimes \alpha}\left( x^{(i)}|x^{(i-1)}\right)
\end{align*}
where $x^{(i)}$ run over vectors in $\cX^{\alpha}$ such that
$T(x^{(i)}_j)=z_i$ for all $j=1,\ldots,\alpha$.
Now we obtain
\begin{align}\label{eq:formula1}
\sum_{z}P_{Z^{n}}(z)^{\alpha} 
& = 
  \sum_{x^{(i)}\in \cC}
    \prod_{j=1}^{\alpha}p(x^{0}_j) \prod_{i=1}^{n} p^{\otimes \alpha}\left( x^{(i)}|x^{(i-1)}\right)
\end{align}
where $\cC = \{s\in\cX^{\alpha}:  T(s_1) = T(s_2)=\ldots= T(s_{\alpha})\}$ is the set of tuples colliding under $T$,
%. Note that  $\cC$ can be written as the disjoint sum $\cC = \bigcup_{t\in\cZ}\cC_t$ of "boxes" 
(illustrated on 
\Cref{fig:1})
\begin{figure}
\begin{tikzpicture}
\draw[pattern=north west lines] (0,0) rectangle node[midway]{$T=t_1$} (1,1);
\draw[pattern=north west lines] (1,1) rectangle node[midway]{$T=t_2$} (3,3);
\draw[pattern=north west lines] (3,3) rectangle node[midway]{$T=t_3$} (4,4);
\node (r) [minimum width=4cm,minimum height=4cm,draw, rectangle] at (2,2) {};
\node[below = 0.1cm of r] {$s_1$};
\node[left =  0.1cm of r] {$s_2$};
\end{tikzpicture}
\caption{States colliding under measurements, for $\alpha=2$.}
\label{fig:1}
\end{figure}
Define accordingly
\begin{align*}
    \mathbf{p}_{\cC}^{\otimes \alpha}(s|s') = \left[ \prod_{j=1}^{\alpha} p(s_j|s'_j) \right]_{s\in\cC,s'\in\cC}
\end{align*}
and 
\begin{align*}
    \mathbf{q}_0 = \left(p(x^0_j)\right)_{x^{0}_j}
\end{align*}
In the matrix-vector notation we have
\begin{align}\label{eq:formula2}
\sum_{z}P_{Z^{n}}(z)^{\alpha} 
& =   \mathbf{q}^T \cdot \left(\mathbf{p}_{\cC}^{\otimes \alpha}\right)^n\cdot  \mathbf{1} 
\end{align}
Since the starting probability $\mathbf{q}$ is positive, the mapping
\begin{align*}
    \| A \| \overset{\text{def}}{=}  \mathbf{q}^T \cdot  |A| \cdot   \mathbf{1}
\end{align*}
where $|A|$ is obtained by applying absolute values to every entry of $A$, is a matrix norm. Since $\mathbf{p}_{\cC}^{\otimes \alpha}$ is nonnegative, we can write
 \Cref{eq:formula2} as
\begin{align*}
    \sum_{z}P_{Z^{n}}(z)^{\alpha}  = \left\| \left(\mathbf{p}_{\cC}^{\otimes \alpha}\right)^n \right\| 
\end{align*}
Now as $n\to\infty$, by Gelfand's formula 
$\|A^n\|^{\frac{1}{n}} = \rho(A)+o(1)$ we obtain
\begin{align*}
\frac{H_{\alpha}(Z^n)}{n} =\frac{1}{1-\alpha} \log
\left(  \left(\sum_{z}P_{Z^{n}}(z)^{\alpha}\right)^{\frac{1}{n}} \right) =  \frac{\rho\left(\mathbf{p}_{\cC}^{\otimes \alpha}\right)}{1-\alpha} + o(1)
\end{align*}
which, after taking the logarithm and dividing by $1-\alpha$, finishes the proof.

To prove the second part, we just replace the Geldand formula by the use of \Cref{lemma:matrix_power_growth}. Then $\mathbf{q}$ doesn't have to be positive.

\section{Simple Noisy Measurements}\label{sec:binary_noise}

Consider the binary Markov chain with transition matrix $M$ observed through a symmetric $\epsilon$-noise binary channel.
Let  $M_{z'z} = \left[p(x,z|x',z')\right]_{x,x'}$ for any fixed $z',z$. Our base matrix is given by
\begin{align*}
  P= 
\begin{bmatrix}
    M_{00} &     M_{01}   \\
    M_{10}   & M_{11}   
\end{bmatrix}  
\end{align*}
The second tensor power equals
\renewcommand{\arraystretch}{1.7} % adds more space between rows
\begin{align*}
P^{\otimes 2} = 
\begin{bmatrix}
    M_{00}  \otimes
    \begin{bmatrix}
     M_{00} & M_{01} \\    
     M_{10} & M_{11}     
    \end{bmatrix}    
    &
    M_{01} \otimes  
    \begin{bmatrix}
     M_{00} & M_{01} \\    
     M_{10} & M_{11}     
    \end{bmatrix}   
    \\
    M_{10} \otimes  
    \begin{bmatrix}
     M_{00} & M_{01} \\    
     M_{10} & M_{11}     
    \end{bmatrix}    
    &
    M_{11}  \otimes 
    \begin{bmatrix}
     M_{00} & M_{01} \\    
     M_{10} & M_{11}     
    \end{bmatrix}    
\end{bmatrix}
\end{align*}
Restricting this to entries $M_{z_1,z_2}\otimes M_{z'_1,z'_2}$
such that $z_1=z'_1$ and $z_2=z'_2$ we obtain
\begin{align}\label{eq:matrix1}
P^{\otimes 2}_{\cC} = 
\begin{bmatrix}
    M_{00}^{\otimes 2} & M_{01}^{\otimes 2} \\
    M_{10}^{\otimes 2} & M_{11}^{\otimes 2} \\
\end{bmatrix}
\end{align}
This matrix is of size $8\times 8$, because every 
submatrix $M_{z_1,z_2}$ is of size $2\times 2$. 
In particular, the spectral radius is a root of an explicit polynomial of degree $8$. 

Let $P_{\epsilon}$ consists of the entries in \eqref{eq:matrix1} 
being of order $O(\epsilon)$. We have
\begin{align*}
    P^{\otimes 2}_{\cC} = P^{0} + P^{\epsilon}
\end{align*}
and by analytic perturbation theory (simple eigenvalue) we know that
\begin{align}\label{eq:perturbation}
     \rho\left( P^{\otimes 2}_{\cC} \right) = \rho\left( P^{0} \right) + O(\epsilon)
\end{align}
We observe that the non-zero entries in $ P^{0}$ are these entries of \eqref{eq:matrix1} that are of the form $M_{z_1,z_2}(x_1,x_2)\cdot M_{z'_1,z'_2}(x'_1,x'_2)$ where 
$z_1=z'_1$, $x_2=x'_2=z_2=z'_2$ (in our case they occur only
in the first and last column). Consider now the square of $P^0$. Observe that the only non-zero entries come from multiplying $P^0$ by the matrix $P'^0$ which restricts the non-zero entries in $P_0$ to
$x_1=x'_1=z_1=z'_1$ and $x_2=x'_2=z_2=z'_2$. We have $(P^0)^k = P^0\cdot (P'^0)^{k-1}$, visualized in our case by
\newcommand\g[0]{\textcolor{gray}{*}}
\begin{align*}
\left(
 \begin{bmatrix}
    \g & \mathbf{0} & \g \\
    \g & \mathbf{0} & \g \\
    \g & \mathbf{0} & \g
\end{bmatrix}\right)^k = 
 \begin{bmatrix}
    \g & \mathbf{0} & \g \\
    \g & \mathbf{0} & \g \\
    \g & \mathbf{0} & \g
\end{bmatrix}
\left(
 \begin{bmatrix}
    \g & \mathbf{0} & \g \\
    \mathbf{0} & \mathbf{0} & \mathbf{0} \\
    \g & \mathbf{0} & \g
\end{bmatrix}\right)^{k-1}
\end{align*}
and therefore
$\rho(P_0) = \rho(P'_0)$. Moreover
\begin{align*}
P_0' =
 \begin{bmatrix}
    p_{00}^2 & \mathbf{0} & p_{10}^2 \\
     \mathbf{0} & \mathbf{0} &  \mathbf{0} \\
    p_{10}^2 & \mathbf{0} & p_{11}^2
\end{bmatrix}
\end{align*}
so we see that non-zero entries of $(P'^0)^k$ are as in the $k$-th power of $M^{\diamond 2}$, for every $k$. Thus
\begin{align}\label{eq:perturbation2}
\rho\left( P^0 \right) = \rho\left( {P'}^{0} \right) = 
\rho\left(M^{\diamond 2} \right)
\end{align}
and combining \Cref{eq:perturbation} with \Cref{eq:perturbation2} finishes the proof. The argument can be extended for other integer $\alpha>1$.

\section{Entropy Rates for Markov Chains}\label{sec:markov_rates}

Let $P$ be the transition matrix of a Markov chain $X_i$,
and measurements $Z_i = T(X_i)$ be given by a deterministic 1-1 mapping $T$. We will obtain formulas for Renyi entropy rates.
Below we illustrate our calculations for $\alpha=2$ and $\cX=2$. The tensored matrx is given by
\renewcommand{\arraystretch}{1.7} % adds more space between rows
\begin{align*}
P^{\otimes 2} = 
\begin{bmatrix}
    p_{00}  
    \begin{bmatrix}
     p_{00} & p_{01} \\    
     p_{10} & p_{11}     
    \end{bmatrix}    
    &
    p_{01}  
    \begin{bmatrix}
     p_{00} & p_{01} \\    
     p_{10} & p_{11}     
    \end{bmatrix}   
    \\
    p_{10}  
    \begin{bmatrix}
     p_{00} & p_{01} \\    
     p_{10} & p_{11}     
    \end{bmatrix}    
    &
    p_{11}  
    \begin{bmatrix}
     p_{00} & p_{01} \\    
     p_{10} & p_{11}     
    \end{bmatrix}    
\end{bmatrix}
\end{align*}
If we restrict this matrix to the entries of the form 
$p_{s_1,s_2}p_{s'_1,s'_2}$ where $s_1=s'_1$ and $s_2=s'_2$ we will obtain
\begin{align*}
P^{\otimes 2}_{\cC} = 
\begin{bmatrix}
     p_{00}^{2} & p_{01}^{2} \\    
     p_{10}^{2} & p_{11}^{2}     
\end{bmatrix}
\end{align*}
as claimed. More generally, for any integer $\alpha>1$
we obtain the matrix with entries 
$p_{s_1,s_2}^\alpha$ which is precisely $P^{\diamond \alpha}$.

\section{Auxiliary Facts on Hiddden Markov Chains}\label{sec:hidden_aux}

Consider a hidden Markov model with the base chain $X_i$ and observations $Z_i$. We will prove that $(X_i,Z_i)$ is a Markov chain, 
with the transition matrix which is the product of the \emph{transition} and \emph{emission} probabilities
\begin{align}\label{eq:joint_chain1}
p(x_i,z_i|x_{i-1},z_{i-1}) = p(x_i|x_{i-1})\cdot p(z_i | x_i).
\end{align}
Indeed, let $I = (x_{i-2},z_{i-2},\ldots,x_1,z_1)$. We have
\begin{align*}
p(x_i,z_i,x_{i-1},z_{i-1}|I) & = p(z_i | x_{i-1},x_{i},z_{i-1},I)\cdot p(z_{i-1} | x_i, x_{i-1},I)\cdot p(x_i,x_{i-1}|I) \nonumber \\
&  =p(z_i|x_i) \cdot p(z_{i-1}|x_{i-1}) \cdot p(x_i,x_{i-1} | I)
\end{align*}
by the output independence property. Also
\begin{align*}
p(x_{i-1},z_{i-1} | I) = p(z_{i-1}| x_{i-1},I)\cdot p(x_{i-1}|I)
\end{align*}
by the output independence property. Dividing the left-hand sides we obtain
\begin{align}\label{eq:joint_chain2}
p(x_i,x_{i-1}|x_{i-1},z_{i-1}) & = \frac{p(x_i,z_i,x_{i-1},z_{i-1}|I)}{p(x_{i-1},z_{i-1} | I)} = p(x_i | x_{i-1},I).
\end{align}
We need to argue that $x_i$ given $x_{i-1}$ is independent of $I$. 
Let $J$ be any subset of $x_k$ where $k> i-2$. We have
\begin{align*}
p(I, J ) & = p(J,x_{i-2},x_{i-3},\ldots,x_1)\cdot p(z_{i-2},z_{i-3},\ldots,z_1 | x_{i-2},x_{i-3},\ldots,x_1,J) \\
& = p(J,x_{i-2},x_{i-3},\ldots,x_1)\cdot \prod_{j=1}^{i-2} p(z_{j} | z_{j-1},\ldots,z_1,x_{i-2},\ldots,x_1,J) \\
& = p(J,x_{i-2},x_{i-3},\ldots,x_1)\cdot \prod_{j=1}^{i-2} p(z_{j} | x_j)
\end{align*}
by the output independence property. Therefore
\begin{align}
p(x_i | x_{i-1},I) & = \frac{p(x_i,x_{i-1},I)}{p(x_{i-1,I})} = 
\frac{p(x_{i},x_{i-1},x_{i-2},x_{i-3},\ldots,x_1)}{p(x_{i-1},x_{i-2},x_{i-3},\ldots,x_1)} \nonumber \\
&= p(x_i | x_{i-1},x_{i-2},\ldots,x_1)
\end{align}
and thus by the Markov property
\begin{align}\label{eq:joint_chain3}
p(x_i | x_{i-1},I)  = p(x_i | x_{i-1}).
\end{align}
Now \Cref{eq:joint_chain2} and \Cref{eq:joint_chain2} imply
\Cref{eq:joint_chain1}.

%\section*{Acknowledgment}


\end{document}

%% file: fig-markovgraph.tex
\definecolor{highlight}{rgb}{0.99,0.78,0.07}
\begin{figure}
\begin{subfigure}[b]{0.5\linewidth}
\begin{tikzpicture}
% These colors should work out nicely when printed in black and white
  [obs/.style={rectangle, font=\sffamily\bfseries}
  ,obsA/.style={obs, fill=blue!120, text=white}
  ,obsB/.style={obs, fill=green!50}
  ]
\node[obsA] (1) at   (90:1.5) {1};
\node[obsB] (2) at (-150:1.5) {2};
\node[obsA] (3) at  (-30:1.5) {3};

\begin{scope}
  [ every path/.style={->, shorten >=1pt} ]
\draw             (1) to[loop above] node[midway,auto]     {0.9} (1);
\draw[bend right] (1) to             node[midway,auto,swap]{0.1} (2);
\draw             (2) to[loop left]  node[midway,auto]     {0.4} (2);
\draw[bend right] (2) to             node[midway,auto,swap]{0.6} (3);
\draw[bend right] (3) to             node[midway,auto,swap]{0.6} (2);
\draw             (3) to[loop right] node[midway,auto]     {0.4} (3);
\end{scope}
\end{tikzpicture}
\caption{The original markov chain.\\{}}
\end{subfigure}
\begin{subfigure}[b]{0.5\linewidth}
\vspace*{-5mm} % optical adjustment due to the hightlight area
\begin{tikzpicture}
  [obs/.style={font=\sffamily\bfseries, rectangle split, rectangle split parts=2,
   rectangle split horizontal}
  ,obsAA/.style={obs,rectangle split part fill={blue!120, blue!120}, text=white}
  ,obsAB/.style={obs,rectangle split part fill={blue!120, green!50}, text=white}
  ,obsBA/.style={obs,rectangle split part fill={green!50, blue!120}}
  ,obsBB/.style={obs,rectangle split part fill={green!50, green!50}}
  ]

\node[obsAA] (11) at  (0,0)   {1 \nodepart{two} 1};
\node[obsAB] (12) at (1,-1)   {1 \nodepart[text=black]{two} 2};
\node[obsBA] (21) at (-1,-1)   {2 \nodepart[text=white]{two} 1};
\node[obsBB] (22) at (0,-2)   {2 \nodepart{two} 2};
\node[obsBA] (23) at (1,-3)   {2 \nodepart[text=white]{two} 3};
\node[obsAB] (32) at (-1,-3)   {3 \nodepart[text=black]{two} 2};
\node[obsAA] (33) at (0,-4)   {3 \nodepart{two} 3};

\node[obsAA] (13) at (2,-2)   {1 \nodepart{two} 3};
\node[obsAA] (31) at (-2,-2)   {3 \nodepart{two} 1};

\begin{scope}[
%  every node/.style={midway, auto, text opacity=0},
%  every path/.style={bend right}
  every path/.style={->, shorten >=1pt}
  ]
\draw (11) to[loop above] (11); % node {0.81}
\draw (11) to[bend left]  (12); % node {0.09}
\draw (11) to[bend right] (21); % node {0.09}
\draw (11) to             (22); % node {0.01}
\draw (12) to[loop right] (12); % node {0.36}
\draw (12) to[bend left]  (13); % node {0.54}
\draw (12) to[bend right] (22); % node {0.04}
\draw (12) to             (23); % node {0.06}
\draw (13) to[loop right] (13); % node {0.36}
\draw (13) to[bend left]  (12); % node {0.54}
\draw (13) to[bend left]  (23); % node {0.04}
\draw (13) to             (22); % node {0.06}
\draw (21) to[loop left]  (21); % node {0.36}
\draw (21) to[bend right] (31); % node {0.54}
\draw (21) to[bend left]  (22); % node {0.04}
\draw (21) to             (32); % node {0.06}
\draw (22) to[loop below] (22); % node {0.16}
\draw (22) to[bend left]  (23); % node {0.24}
\draw (22) to[bend right] (32); % node {0.24}
\draw (22) to             (33); % node {0.16}
\draw (23) to[loop right] (23); % node {0.16}
\draw (23) to[bend left]  (22); % node {0.24}
\draw (23) to[bend left]  (33); % node {0.24}
\draw (23) to             (32); % node {0.16}
\draw (31) to[loop left]  (31); % node {0.36}
\draw (31) to[bend right] (21); % node {0.54}
\draw (31) to[bend right] (32); % node {0.04}
\draw (31) to             (22); % node {0.06}
\draw (32) to[loop left]  (32); % node {0.16}
\draw (32) to[bend right] (22); % node {0.24}
\draw (32) to[bend right] (33); % node {0.24}
\draw (32) to             (23); % node {0.16}
\draw (33) to[loop below] (33); % node {0.16}
\draw (33) to[bend right] (32); % node {0.24}
\draw (33) to[bend left]  (23); % node {0.24}
\draw (33) to             (22); % node {0.16}
\end{scope}

\begin{scope}[on background layer]
\path[fill=highlight, draw=highlight, line width=8pt, on background layer]
 plot [smooth cycle, tension=1] coordinates {%
 (11.north east)
 (22.north east)
 (13.north east)
 (13.south east)
 (22.south east)
 (33.south east)
 (33.south west)
 (22.south west)
 (31.south west)
 (31.north west)
 (22.north west)
 (11.north west)
 };
\end{scope}

\end{tikzpicture}
\caption{The second tensor.\\The set $\cC$ of colliding states is highlighted.}
\end{subfigure}
\caption{A three state hidden markov chain, the color is observable.}
\label{fig:markovgraph}
\end{figure}